\documentclass[twocolumn]{aastex631}
\usepackage[percent]{overpic}

\shorttitle{Eruptive and confined flares}
\shortauthors{Maity et al.}

\begin{document}

\title{Photospheric Lorentz force changes in eruptive and confined solar flares}

\author[0009-0005-4347-9044]{Samriddhi Sankar Maity}
\affiliation{Indian Institute of Astrophysics, Bangalore, India}
\affiliation{Joint Astronomy Programme and Department of Physics, Indian Institute of Science, Bangalore, India}

\author[0000-0001-6457-5207]{Ranadeep Sarkar}
\affiliation{University of Helsinki, Finland}

\author[0000-0002-0181-2495]{Piyali Chatterjee}
\affiliation{Indian Institute of Astrophysics, Bangalore, India}

\author[0000-0002-0452-5838]{Nandita Srivastava}
\affiliation{Udaipur Solar Observatory, Physical Research Laboratory, Udaipur, India}

\begin{abstract}

Solar flares are known to leave imprints on the magnetic field at the photosphere, often manifested as an abrupt and permanent change in the downward-directed Lorentz force in localized areas inside the active region. Our study aims to differentiate eruptive and confined solar flares based on the vertical Lorentz force variations. We select 26 eruptive and 11 confined major solar flares (stronger than the GOES M5 class) observed during 2011-2017. We analyze these flaring regions using SHARP vector-magnetograms obtained from the NASA's Helioseismic and Magnetic Imager (HMI). We also compare data corresponding to 2 synthetic flares from a $\delta$--sunspot simulation reported in Chatterjee et al. [Phys. Rev. Lett. 116, 101101 (2016)]. We estimate the change in the horizontal magnetic field and the total Lorentz force integrated over an area around the polarity inversion line (PIL) that encompasses the location of the flare. Our results indicate a rapid increase of the horizontal magnetic field along the flaring PIL, accompanied by a significant change in the downward-directed Lorentz force in the same vicinity. Notably, we find that all the confined events under study exhibit a total change in Lorentz force of $< 1.8 \times 10^{22}$ dyne. This threshold plays an important factor in effectively distinguishing eruptive and confined flares. Further, our analysis suggests that the change in total Lorentz force also depends on the reconnection height in the solar corona during the associated flare onset. The results provide significant implications for understanding the flare-related upward impulse transmission for the associated coronal mass ejection.

\end{abstract}

\section{Introduction}

Solar flares and Coronal Mass Ejections (CMEs) are considered the two most violent and energetic phenomena occurring in the solar atmosphere due to the sudden release of energy.  A typical flare is recognized by a quick increase in light emission in a broad range of the electromagnetic spectrum that affects the solar atmosphere, while a CME consists of plasma and high-energy particles that are expelled from the Sun. They are responsible for significant space weather impacts on Earth \citep{siscoe2000spwea, daglis2004spwea, chen2017fluxropes, green2018solar}. Therefore, understanding the source region characteristics of these energetic solar events has become one of the most important goals of space science research.
A solar flare in relation to CMEs is classified into two types: eruptive and confined \citep{moore2001onset}. Eruptive flares are associated with CMEs, while confined flares do not have associated CMEs.

Previous observations have shown that flares and CMEs are different manifestations of the same energy-release process \citep{harrison2003association,zhang2001temporal}. Moreover, \cite{zhang2001temporal} have shown that the phase of rapid acceleration of CMEs in the inner corona is temporarily correlated with the rise time of the associated soft X-ray flares. In spite of the intrinsic correlation between flares and CMEs, observations have shown that not all flares are associated with CMEs \citep{andrew2003association,yashiro2005visibility,Yashiro2009Flares,Webb2012CMEObservations,Youssef2012Flare}. Active regions (ARs) with complex topology are the primary sources of large flares and most energetic CMEs \citep{zirin1987greatflares,sammis2000dependence,Yang2017Flare}. During a flaring event, the magnetic field reorganizes rapidly in the corona owing to the low Alfv\'en travel time, leading to the eruption of magnetic flux rope and the subsequent formation of post-flare loops beneath the current sheet, in accordance with the standard flare model \citep{Carmichael1964Flares,Hirayama1974Flares,Sturrock1966Flares,KoppPneuman1976Flares,shibata2011flares}. The flaring process converts the magnetic free energy into kinetic, thermal energy and non-thermal energy that accelerates particles. Although the photosphere is much denser than the corona, the photospheric magnetic field can respond to sudden coronal restructuring during flaring events \citep{wang2015magfields,aulanier2016tail,toriumi2019solarastrophy}. Counter-intuitively, observations have shown that the photospheric magnetic field does experience significant changes during flares \citep{wang2015magfields,toriumi2019solarastrophy}. Since there is no practical or direct method to measure the vector magnetic field in the coronal volume, it is challenging to quantitatively investigate the temporal evolution of non-potential parameters (e.g., magnetic free energy) \citep{wiegelmann2014solaratmos}. Therefore, the temporal or spatial evolution of parameters in the source region that can only be estimated from the photospheric \citep{petrie2010abrupt} and chromospheric \citep{Kleint2017Chromospheric} magnetic field (e.g., the change in the net Lorentz force) becomes a major probe to study the changes associated with the flare. \cite{hudson2008flare} were the first to suggest that photospheric magnetic fields should become more horizontal after the flare due to the effect of vertical Lorentz forces on the solar surface.  Developing this model further, \cite{fisher2012global} gave a practical approach to calculate the net Lorentz force acting on the solar photosphere. They found an increase in the horizontal magnetic field ($B_h$), particularly around the polarity inversion line (PIL), and an associated large and abrupt downward change in the vertical Lorentz force. Previous studies have also found that large eruptive flares are associated with a sudden downward change in the Lorentz force \citep{petrie2010abrupt,petrie2012abrupt}. In contrast to $B_h$, the vertical magnetic field ($B_z$) varies much less during the flare without a clear pattern \citep{sun2017magnetic}.
On the other hand, the sunspot area weighted horizontal gradient of the vertical magnetic field is found to follow a distinct pattern before a flare, providing potential predictive capability \citep{korsos2015flare}. This behavior of the weighted gradient is also marked by the approaching–receding motion of the barycenters of opposite polarities before the flare. \cite{sarkar2018comparative} compared the magnitude of changes in the horizontal magnetic field ($B_h$) and the net vertical Lorentz force associated with eruptive and confined flares occurred in a same active region (AR). They reported that the flare associated changes in magnetic parameters are larger for eruptive flares than for confined ones. Extending the study to large recurrent flares, \cite{Sarkar_2019} found that the change in net vertical Lorentz force acts as an excellent proxy to predict the recurrent large flaring events from a same AR. \cite{vasantharaju2022magimprint} reported that the vertical Lorentz force changes during flares near PILs correlate well with the flare strength. However, no clear classification of the association of flares with CMEs has been made to statistically distinguish them by the net change in Lorentz force.

We now have evidence of rapid and permanent changes in the longitudinal and transverse magnetic fields linked to solar flares in the photosphere thanks to the availability of high-cadence photospheric vector magnetograms from the ground- and space-based telescopes \citep{sudol2005longitudinal,petrie2010abrupt,wang2012response,gosain2012evidence,sun2017magnetic,castellanos2018statistical,petrie2019abrupt,liu2022horizontal,Kazachenko2022Variability,Kazachenko2023Database}.

The magnetic implosion conjecture is frequently used to explain observational evidence of the rise in the horizontal component of the magnetic field in the solar atmosphere \citep{hundson2000implosions}. It states that in a low plasma $\beta$ environment, the coronal loops must contract during a transient event, such as a flare or a CME, in order to lower the magnetic energy. According to \cite{hudson2008flare} and \cite{fisher2012global}, the release of free magnetic energy should be accompanied by a decrease in the magnetic pressure and volume. A MHD wave that propagates downward towards the photosphere and perturbs the field there may also be excited by this abrupt change in the corona which increase the horizontal component of the magnetic field in the photosphere near the polarity inversion line (PIL) \citep{fletcher2008impulsive,hudson2008flare,wheatland2018photospheric}. \cite{li2011comparison} discovered that following the flare, the horizontal magnetic field close to the flaring PIL shows a change in both observation and simulation.

In this study, we analyzed the photospheric field variations of 37 events by using the 12-minute cadence vector magnetogram from the Helioseismic and Magnetic Imager (HMI) onboard Solar Dynamics Observatory (SDO). We further augment the event list with the addition of the two of four synthetic flares that occurred in the 3-dimensional magneto-hydrodynamic (MHD) simulation reported in \citep{chatterjee2016repeatedflare} and \citep{korsos2018weighted}. The primary motivation of this work is to understand the correlation of the change in vertical Lorentz force with the eruptivity of the flares and uniquely classify flares associated with CMEs in terms of the change in Lorentz force. The rest of the paper is structured as follows. In Section 2, we describe the data set and methods employed. The results are presented in Section 3. Finally, we discuss the results and summarize our conclusions in Section 4.

\begin{figure*}[!t]
\centering
\includegraphics[width=\textwidth]{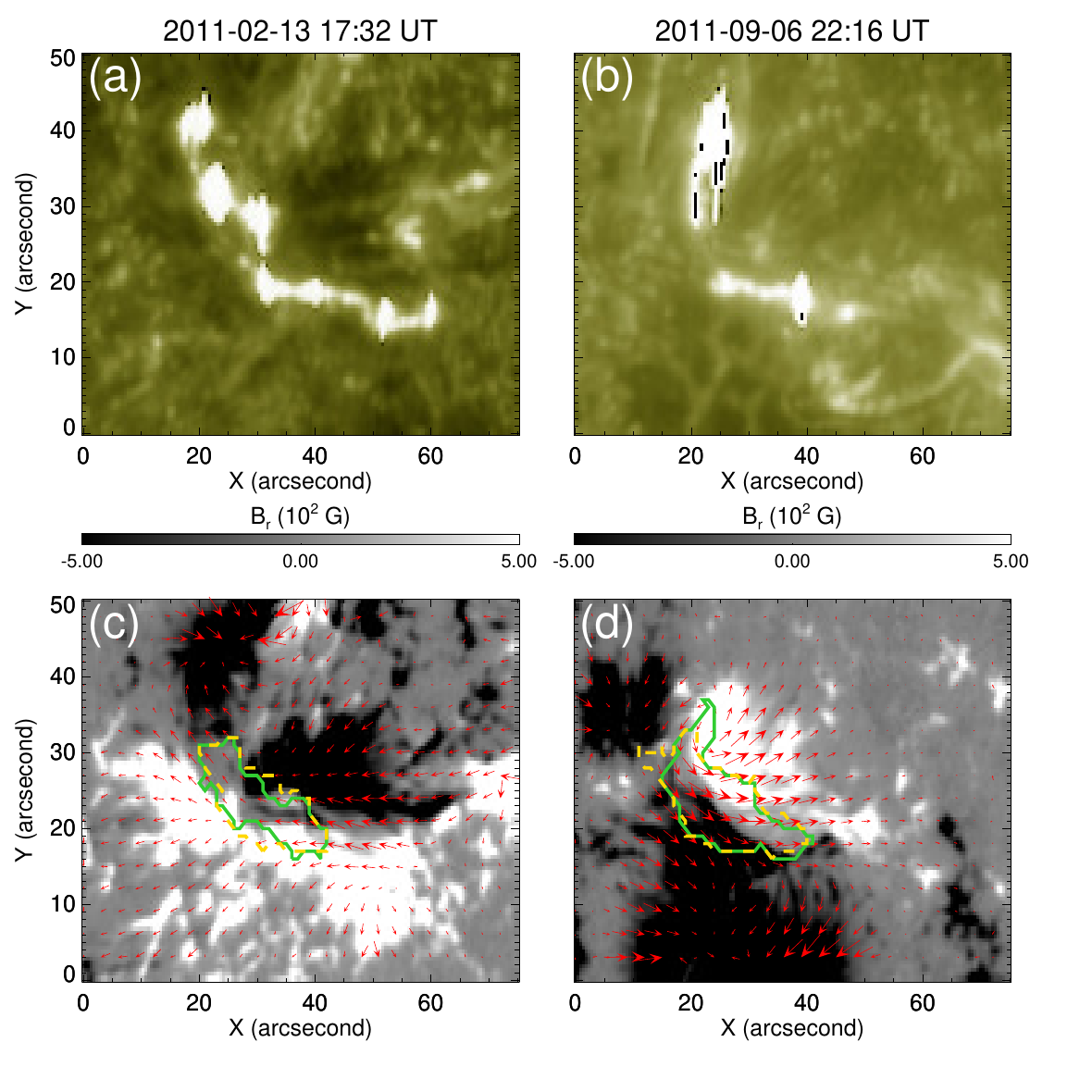}
\caption{Illustrations of two eruptive events to identify the regions of interest (RoIs) of the magnetic imprints (MIs). The panels a and b are AIA 1600\AA\ images of the two flaring events occurred on 2011 February 13 at 17:38 UT and 2011 September 06 at 22:20 UT respectively. The panels c and d represent the radial magnetic field $B_r$ whose strength is indicated by colorbars. The horizontal component of the magnetic field is shown by the red arrows. The over-plotted contours mark the RoIs selected based on the individual difference maps of the $B_h$ (yellow dashed line) and $F_r$ (solid green lines).}
\label{fig:eruptive_events}
\end{figure*}

\begin{figure*}[!ht]
\centering
\includegraphics[width=\textwidth]{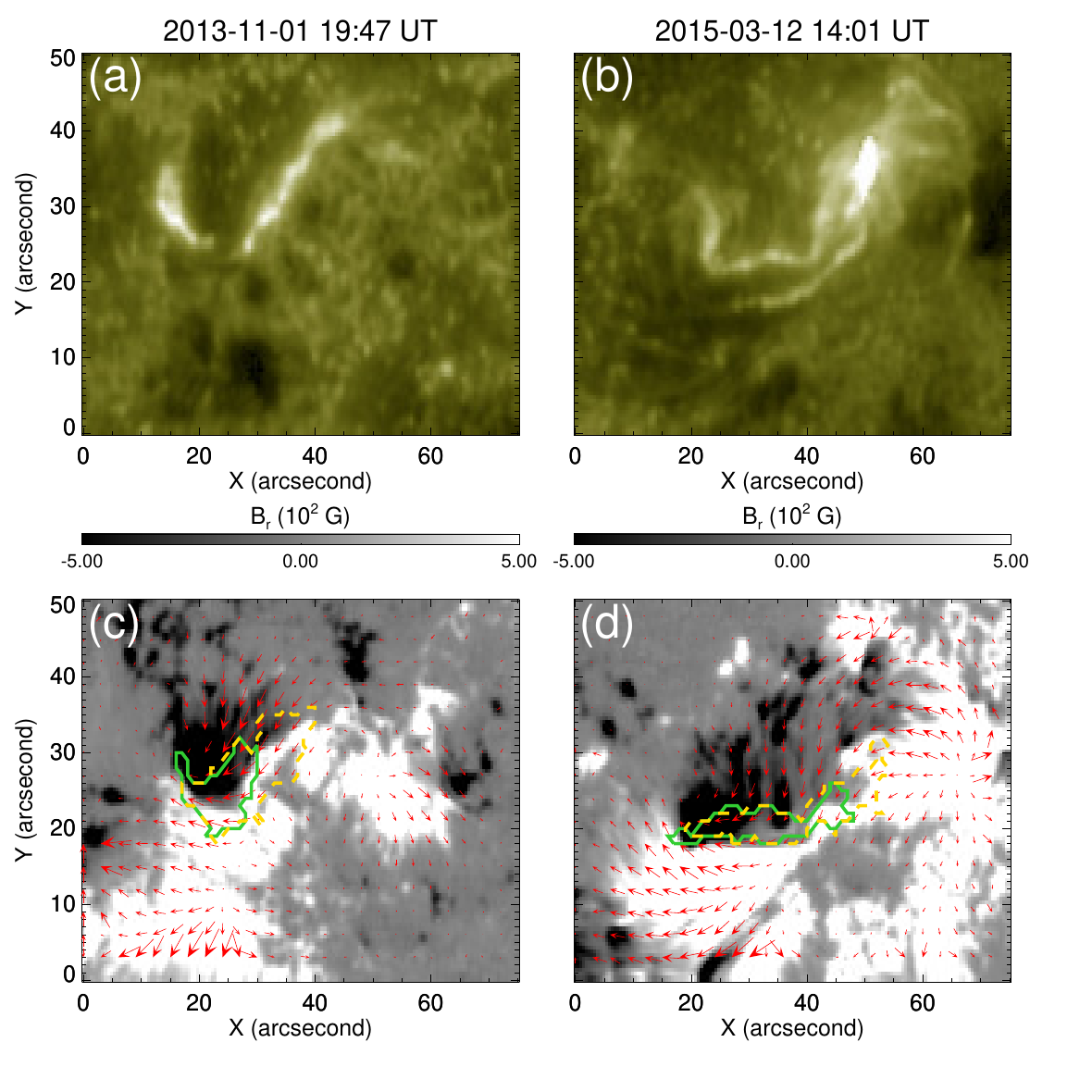}
\caption{Similar to figure \ref{fig:eruptive_events} but for confined events occurred on (a), (c) 2013 November 01 at 19:53 UT and (b), (d) 2015 March 12 at 14:08 UT.}
\label{fig:confined_events}
\end{figure*}

\section{Data and Methods}

\subsection{Observational data}
Based on the event catalog provided by \citet{jing2018statistical}, we selected 37 major solar flares including 15 X- and 22 M-class flares originated from 26 active regions (ARs) located within $\pm 45^{\circ}$ of the solar disk center. The selected events comprise both eruptive and confined flares over a seven-year period from 26 January 2011 to 11 December 2017. For each event, we used the vector magnetograms of the AR during the flaring event obtained by HMI \citep{schou2012design}, on board SDO \citep{pesnell2012solar}. In particular, we used the HMI vector magnetogram series from the version of Space Weather HMI Active Region Patches (SHARP, \cite{turmon2010active} having a spatial resolution of 0.5 arcsecond with a temporal cadence of 12 minutes. HMI measures the Stokes parameters at six wavelengths centered on Fe I 6173 \AA\ absorption line with a bandwidth of 76 \AA. Based on these observations, the photospheric vector magnetic field is derived by inverting full set of Stokes parameters using the Milne-Eddington inversion approach \citep{borrero2011very} to obtain the vector magnetic-field components in the photosphere. As part of the additional data pre-processing steps, a coordinate transformation is performed to remap the vector fields onto the Lambert cylindrical equal area projection. Subsequently, the components of the vector field are converted into Heliocentric spherical coordinates ($B_r, B_{\theta}, B_{\phi}$). Apart from the 12 minutes cadence, HMI also has high cadence vector magnetic field data with temporal cadence of 135 seconds. However, we carried out our analysis using the 12 minutes cadence vector magnetogram data due to its lower noise level than that of the 135 seconds cadence for full disk vector magnetogram data \citep{sun2017magnetic}. Moreover, the flare related field changes are sufficiently captured with the cadence of 12 minutes as studied previously \citep{sarkar2018comparative,Sarkar_2019}

We have also used the 1600 \AA\ images provided by AIA \citep{lemen2012atmospheric} on board SDO to approximate the location of flare ribbons, which helps identify and select the Region of Interest on the vector magnetograms. In order to characterize and analyze the evolution of magnetic field changes, we chose a 12-hour time-window around the time of the solar flare, encompassing six hours both before and after the peak of the flare. The flare start, peak and end time is determined from the Hinode catalog \footnote{https://hinode.isee.nagoya-u.ac.jp/flare\_catalogue/}.

\subsection{Simulation data}
In order to compare the observations with simulation, we focus on the numerical case study presented in \cite{chatterjee2016repeatedflare} for our analysis. We provide a concise overview of the model setup here for completeness. The box-shaped computational domain has horizontal extents of -18 $\mathrm{Mm}<\mathit{x},\mathit{y}<$ 18 $\mathrm{Mm}$ and a vertical one of -8.5 $\mathrm{Mm}<\mathit{z} <$ 16.5 $\mathrm{ Mm}$, with uniformly spaced grid with d$\mathit{x}$ = d$\mathit{y}$ = 96 km and d$\mathit{z}=48$ km, rotating with a angular velocity similar to Sun, forming an angle of $30 ^{\circ}$ with the vertical z-direction. A constant gravity, $g_z$, points in the negative $z$-direction. For the calculation, authors use the Pencil Code\footnote{https://github.com/pencil-code} \citep{brandenburg21} - a fully compressible higher-order finite difference tool. Beginning from the initial state, the simulation was run for 263 minutes of solar time. The initial subsurface horizontal magnetic sheet breaks up, rises, and emerges through the surface like a newly emerging AR after about 145 minutes. There were four flaring eruptions recognized. The first two flares in the simulation are B and C-class have onset times at 167.5 and 197.2 minutes, respectively and are analyzed for this work. We excluded the other two flares from the analysis due to the presence of numerical artifacts.

The flares reported in the above work released energies of $\approx 2\times10^{31}$\,ergs commensurate with B- and C-class flares. First of all, note that it is computationally challenging to produce solar flare energies of M and X class in solar MHD simulations with photospheric flux emergence due to requirement of very high magnetic Reynolds number, domain size and the wall clock time for which such simulations can be run. Conversely, analyzing stronger flares proves easier in observations, while changes caused by weaker flares might go undetected by current instruments. \cite{vasantharaju2022magimprint,Kumar2020sesmic}, incorporated C-class flares in their study, despite the uncertainty in the vector magnetic field data associated with them. Consequently, we decided to forego additional observational analysis of weak flares, focusing instead on understanding the magnetic imprint problem through numerical simulation. Our approach involves tackling the issue from two contrasting perspectives of variability. We intend to comprehensively understand the problem by examining minor flares through numerical simulation and major flares through observational data.

\begin{figure*}[!t]
\centering
\includegraphics[scale=0.56]{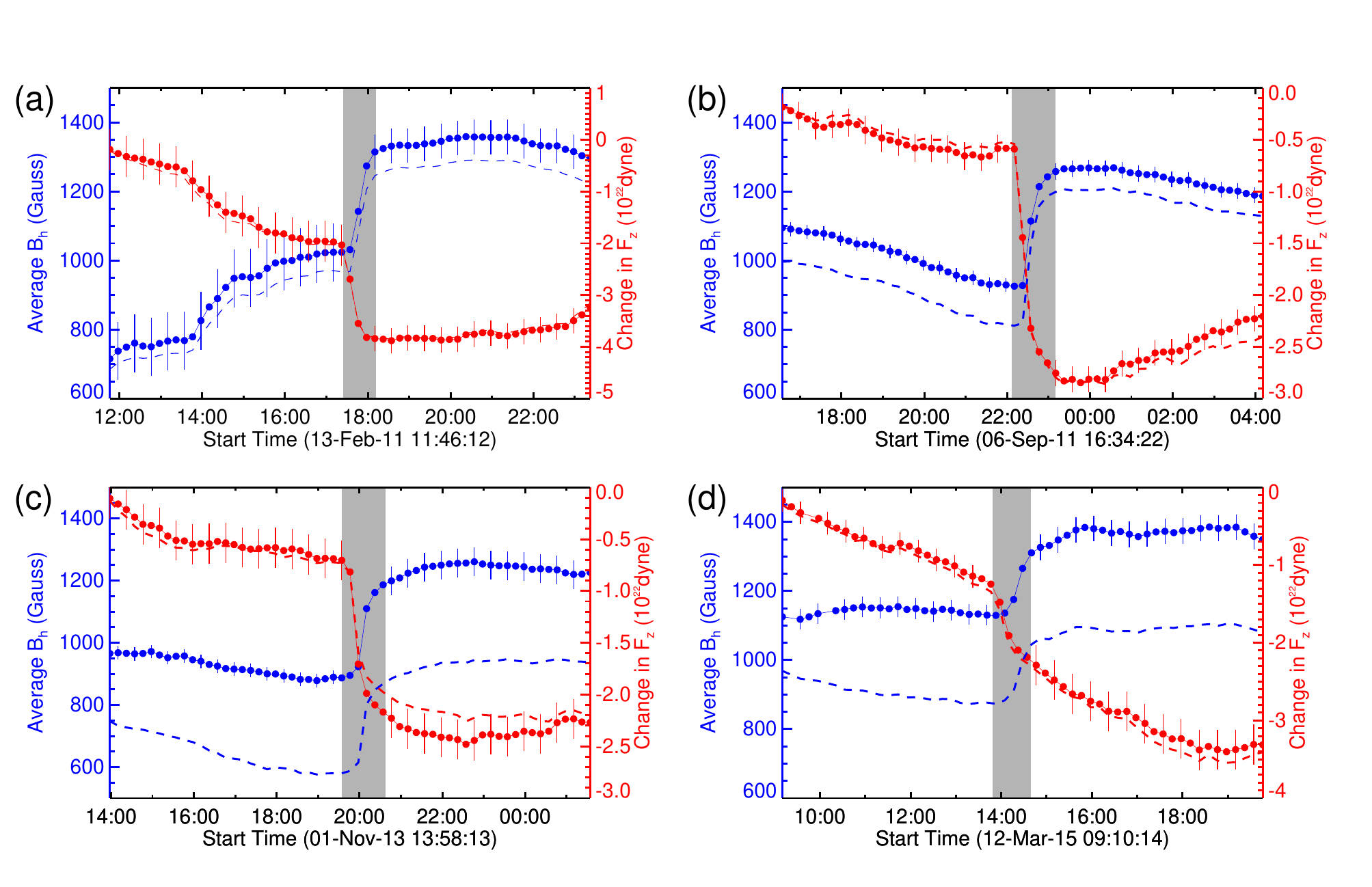}
\caption{Temporal evolution of average horizontal magnetic fields (blue) and vertical Lorentz force (red) calculated by Lorentz force (solid) and horizontal magnetic field (dashed) contouring method. The panels a and b represent eruptive, whereas the panels c and d represent confined events. The shaded region corresponds to the duration of field change. The vertical error bars represents the fluctuations at a 3$\sigma$ level in both pre- and post flare states.}
\label{fig:Obs_temporal}
\end{figure*}

\subsection{Lorentz force and Masking Algorithm}
We utilized the formulation proposed by \cite{fisher2012global} to calculate the total changes in the Lorentz force. The variation in the horizontal and vertical component of the Lorentz force within a time interval of $\delta t$ is computed using the following equations.

\begin{equation}
\label{eqn:Fz}
\delta F_z = \frac{1}{8 \pi} \int_{A_{ph}} (\delta B_h^2 - \delta B_z^2) \,dA
\end{equation}
\begin{equation}
\label{eqn:Fh}
\delta F_h = - \frac{1}{4 \pi} \int_{A_{ph}} \delta (B_h B_z) \,dA
\end{equation}

Here $B_h$ and $B_z$ represent the horizontal and vertical components of the magnetic field, respectively, and $F_h$ and $F_z$ are the horizontal and vertical components of the Lorentz force calculated over the volume of the Active Region (AR). The domain $A_{ph}$ corresponds to the photospheric area containing the AR, and $dA$ is the elementary surface area on the photosphere. Similar to the approach  by \cite{petrie2012abrupt}, we have reversed the signs in equations (\ref{eqn:Fz}) and (\ref{eqn:Fh}) in comparison to Equations (9) and (10) of \cite{fisher2012global}. This change accounts for considering the equal and opposite forces acting on the above atmosphere from below. Since significant changes in a horizontal magnetic field and Lorentz forces associated with flares are expected near the Polarity Inversion Line (PIL) \citep{wang2006rapid,petrie2010abrupt,petrie2012abrupt,sarkar2018comparative,Sarkar_2019}, we focused our analysis on subdomains near the PIL within the flare productive region of each AR. The reason behind this is based on the assumption that the magnetic field on side boundaries enclosing these subdomains remains relatively constant over time. Additionally, we consider the magnetic field strength on the top boundary to be negligible compared to that at the lower boundary of the photosphere. Consequently, in equation (\ref{eqn:Fz}) and (\ref{eqn:Fh}), only changes in the photosphere magnetic field contribute to the surface integrals, allowing us to estimate the net change in the Lorentz force acting on the photosphere from the volume above the atmosphere.

We developed a semi-automatic tool to select the sub-region in which we analyzed the flare-associated Lorentz force changes for all the events. As the significant Lorentz force changes are expected to happen very close to the flare peak time, we select time frames approximately 30 minutes before and after the peak phase of the associated flare. Next, we create difference maps of the horizontal magnetic field and vertical Lorentz force estimated at the above-mentioned time frames separately. We use information of both the horizontal magnetic field ($B_h$) and vertical Lorentz force ($F_r$) independently to avoid any selection bias in identifying the areas where the most significant changes occurred as discussed in \cite{vasantharaju2022magimprint,liu2022horizontal,yadav2022statistical,petrie2012abrupt}. To these difference maps, we apply a threshold to select the sub-regions that demarcate the area of positive change ($>$ 100 Gauss) in the horizontal magnetic field or the negative change ($< - 10^{19}$ dyne) in the vertical component of the Lorentz force. Applying this method, we find several sub-regions within the AR. In order to find the correct region of interest (RoI), we manually select the sub-region which is in the closest proximity to the flaring location observed in the AIA images. Figure \ref{fig:eruptive_events} and \ref{fig:confined_events} illustrates the identified RoIs based on both the $B_h$ and $F_r$ difference maps for the two eruptive and confined events from Table (\ref{tab:events}) respectively.

\begin{figure*}[!t]
\centering
    \includegraphics[width=0.8\textwidth]{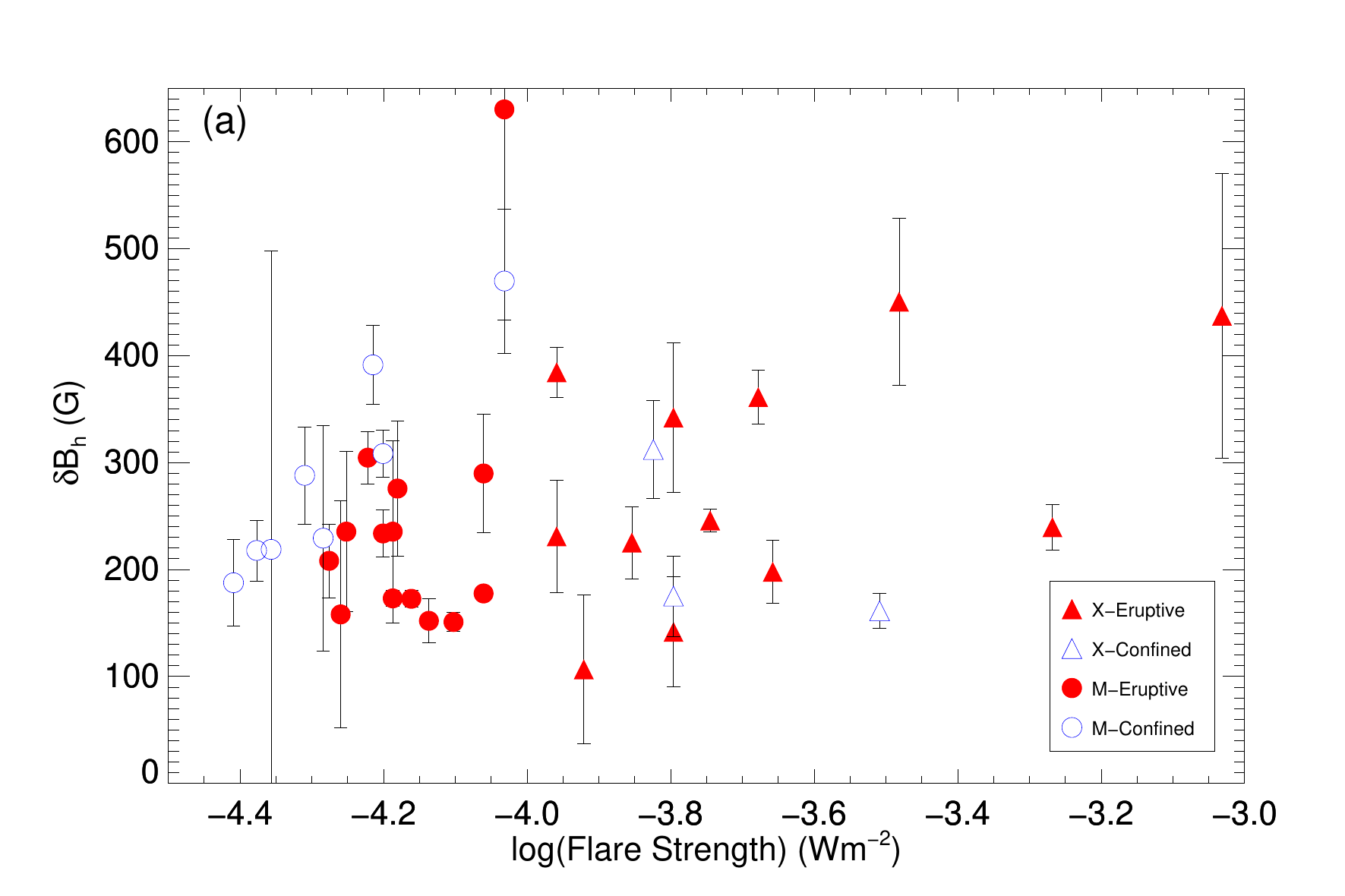}
    \includegraphics[width=0.8\textwidth]{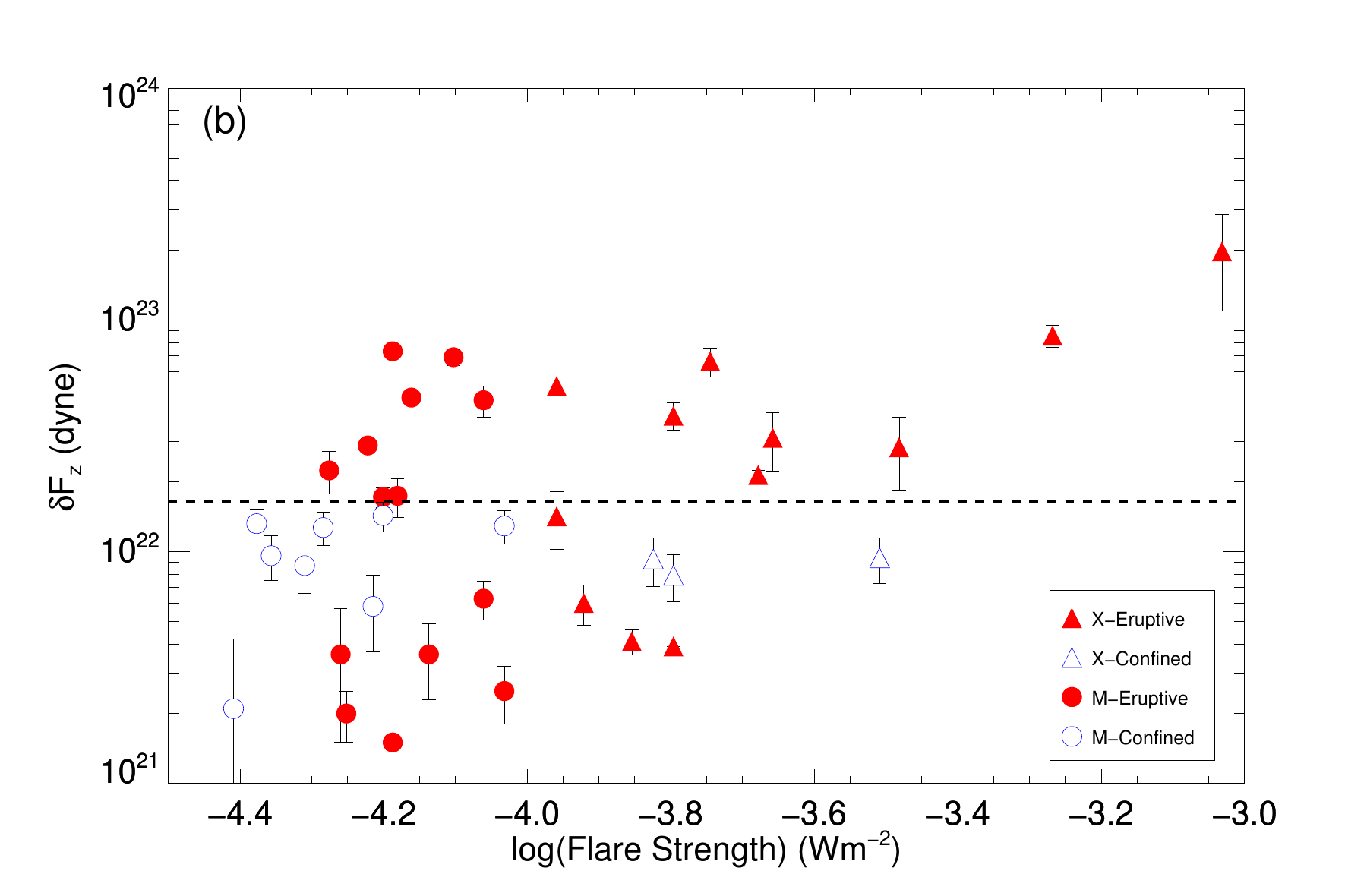}
\caption{Scatter plot of (a) Average horizontal magnetic field change $\delta B_h$ vs logarithmic flare strength and (b) Vertical Lorentz force change $\delta F_z$ vs logarithmic flare strength for RoIs identified based on the difference maps of horizontal magnetic field. Filled and empty symbols correspond to the eruptive and confined flares, respectively. The triangular and circular symbols are for X-class and M-class flares, respectively. The horizontal dashed line indicates the threshold Lorentz force above which no confined flares are observed.}
\label{fig:bhcontour_plot}
\end{figure*}

\section{Results \& Discussion}
The characteristic variations in the average horizontal magnetic field and the total downward Lorentz force for two eruptive and two confined flaring events which are observed on February 13, 2011 (Event No 1); September 06, 2011 (Event No 7) and November 1, 2013 (Event No 16) and March 12, 2015 (event No 33) are described here as examples. Then, for all 37 events, we summaries their variations with flare strength.

\begin{figure*}[!t]
\centering
    \includegraphics[width=0.8\textwidth]{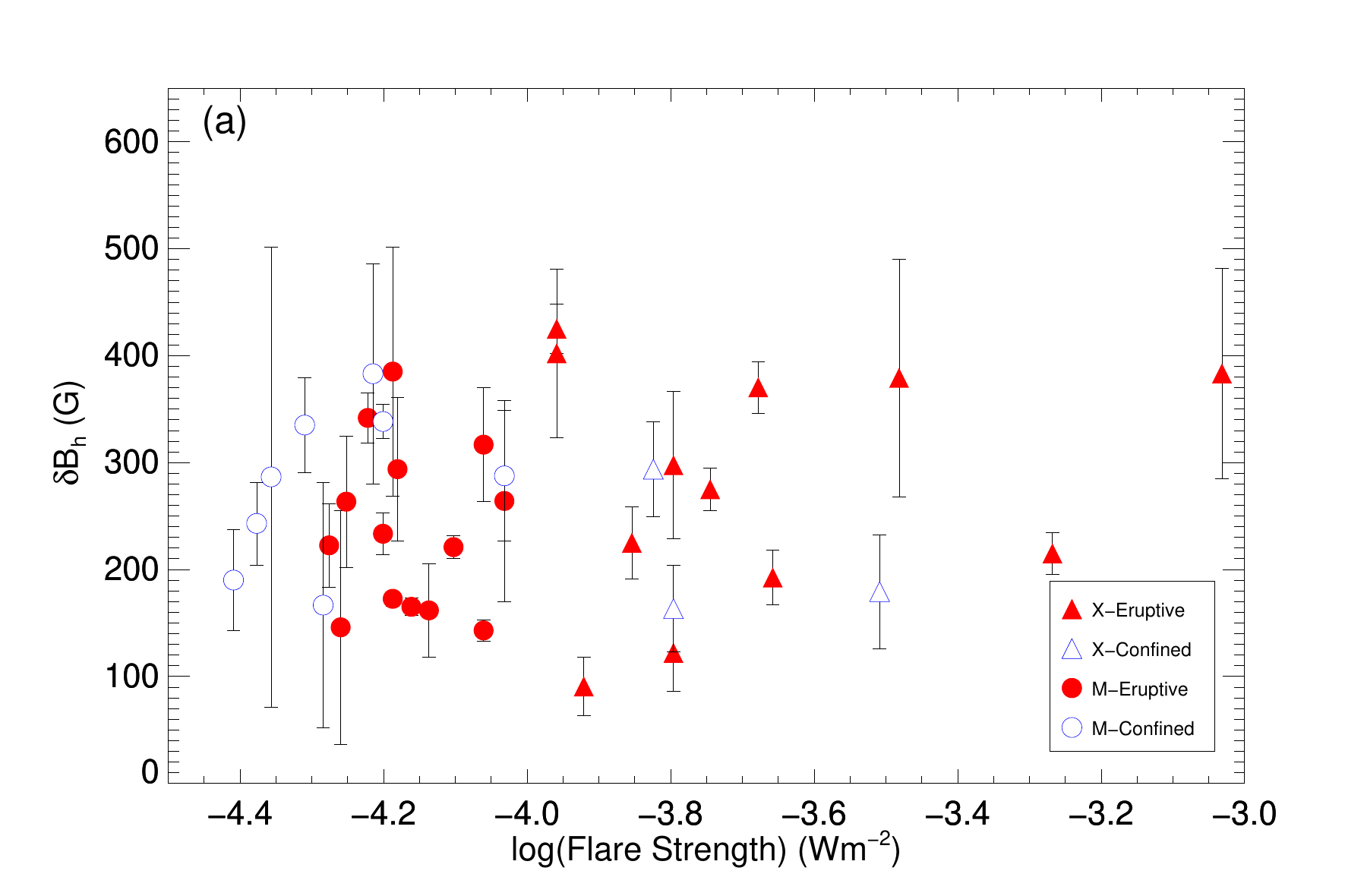}
    \includegraphics[width=0.8\textwidth]{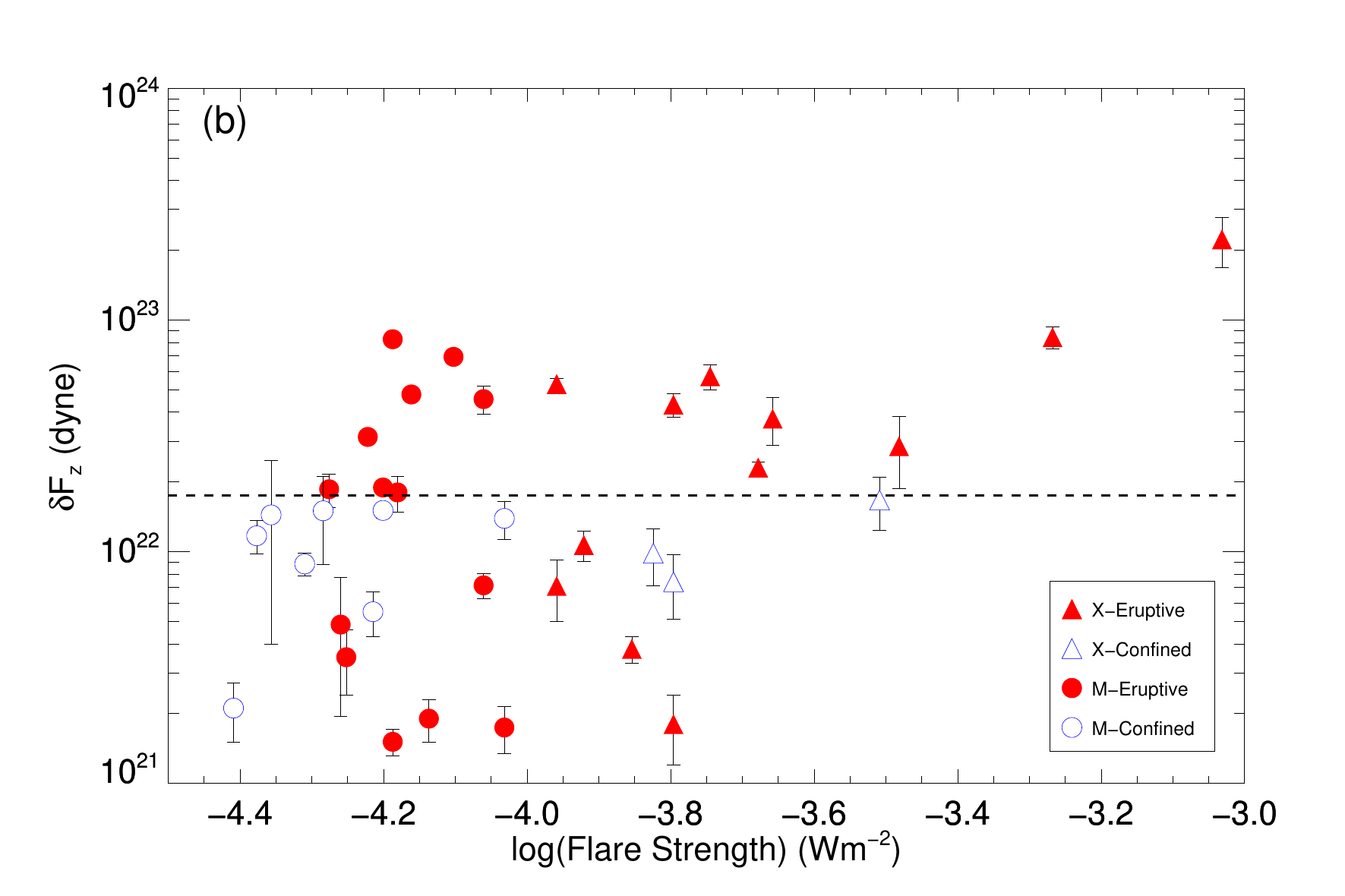}
\caption{Similar to Figure \ref{fig:bhcontour_plot} but for RoIs identified based on the difference maps of Lorentz force.}
\label{fig:flcontour_plot}
\end{figure*}

\begin{figure*}[!t]
\centering
\begin{overpic}[width=0.8\linewidth]{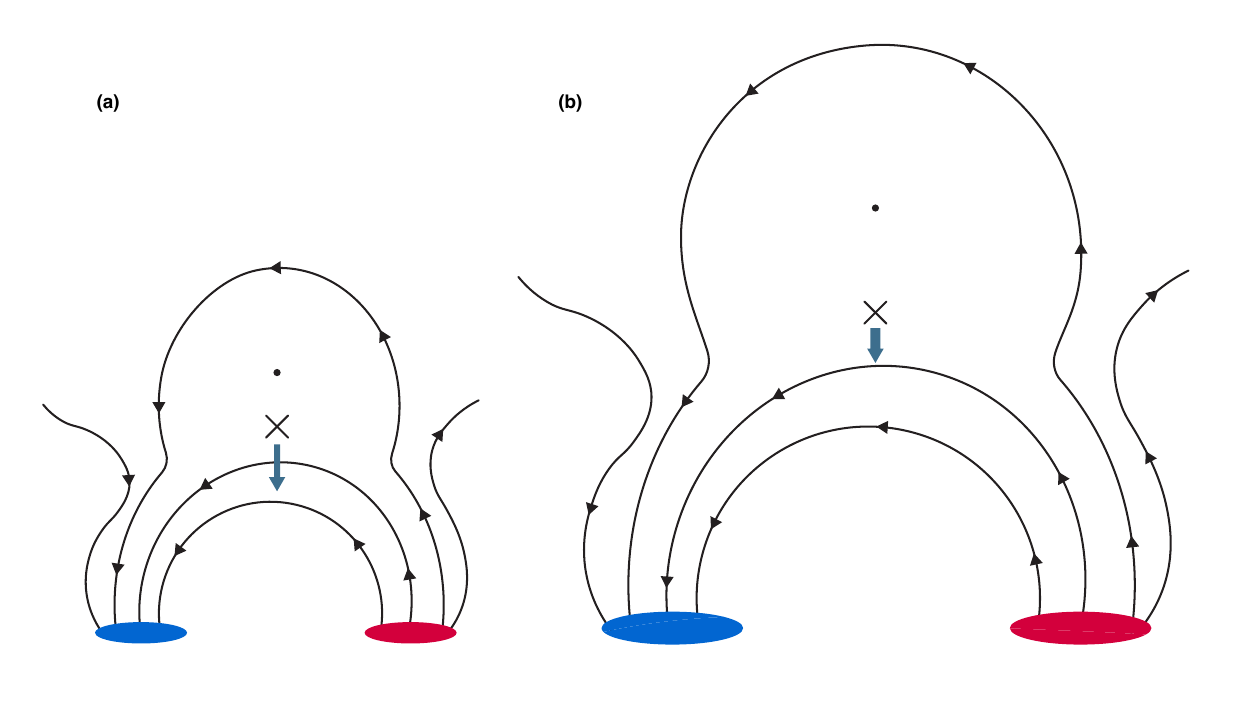}
\put (20,57 ) {(a)}
\put (68,57) {(b)}
\end{overpic}
\caption{A sketch of the magnetic field configuration of (a) eruptive event with higher Lorentz force change and (b) eruptive event with lower Lorentz force change. Red and blue filled regions are positive and negative polarity regions whereas solid lines refer to the magnetic field lines. The X mark represents the location of the reconnection site and the downward arrow implies the direction of vertical Lorentz force responsible for the increase of horizontal magnetic field.}
  \label{fig:Cartoon}
\end{figure*}

\subsection{Evolution of $B_h$ and $F_z$}
After successful identification of the flaring region using two different methods based on the $B_h$ and $F_z$ difference maps, we continue our analysis within that sub-region. We studied the temporal evolution of the average $B_h$, and the total downward Lorentz force change, $F_z$, in the selected region near the PIL for each case. As an example, the time variation of average $B_h$ and downward $F_z$ for the same two eruptive (Event No 1 and 7) and confined events (Event No 16 and 33) are shown in Figure \ref{fig:Obs_temporal}. The top and bottom panels represent the variation of average $B_h$ and downward $F_z$ over time for each eruptive and confined events, respectively. All flaring events show a abrupt change in both $B_h$ and $F_z$ using both the methods. The shaded region in these plots indicates the field change duration. The error bars depict fluctuations corresponding to a 3$\sigma$ significance in both pre and post flaring states. These fluctuations are determined through separate linear regression of the temporal changes in $B_h$ and $F_z$ before and after the shaded time interval. The error analysis is performed using a time window of 6 hours and a resolution of 12 minutes in each states. The errors specified in columns 8-11 of Table (\ref{tab:events}) are determined by averaging the errors from the pre-flare and post-flare states.

Within the RoI determined from the $F_z$ difference maps, the average change in $B_h$ and $F_z$ for eruptive events as shown in Figure \ref{fig:Obs_temporal} found to be $293.8$ Gauss, $1.8 \times 10^{22}$ dyne for Event ID 1 and $370.4$ Gauss, $2.3 \times 10^{22}$ dyne for Event ID 7. Similarly, for confined events, the average change in $B_h$ and $F_z$ in the RoI given by the Lorentz force change were $33$8 Gauss, $1.5 \times 10^{22}$ dyne for Event ID 16 and $243$ Gauss, $1.2 \times 10^{22}$ dyne for Event ID 33.

On the other hand, when the RoI was identified based on the $B_h$ difference map, the average changes in $B_h$ and $F_z$ for eruptive events were $275.1$ Gauss, $1.7 \times 10^{22}$ dyne for Event ID 1 and $361$ Gauss and $2.1 \times 10^{22}$ dyne for Event ID 7, respectively. For confined events, the average changes in $B_h$ and $F_z$ in the RoI given by the change of horizontal magnetic field were $308.3$ Gauss, $1.4 \times 10^{22}$ dyne for Event ID 16 and $217.2$ Gauss, $1.3 \times 10^{22}$ for Event ID 33.

Figure \ref{fig:Obs_temporal} shows that the enhancement of average $B_h$ is permanent throughout the post-flare phase (at least within 6 hours of time window after the flare peak time), which agrees with the previous studies \citep{wang2012relationship,petrie2012abrupt,sun2017magnetic,liu2022horizontal}. The total downward $F_z$ is observed to show an abrupt decrease during the flare interval, which is also irreversible. The average changes in $B_h$ and the total change in vertical $F_z$ for all 37 events, analyzed for different RoI identification methods, are tabulated in Table (\ref{tab:events}).

\begin{figure*}[!t]
\centering
\includegraphics[width=0.8\textwidth]{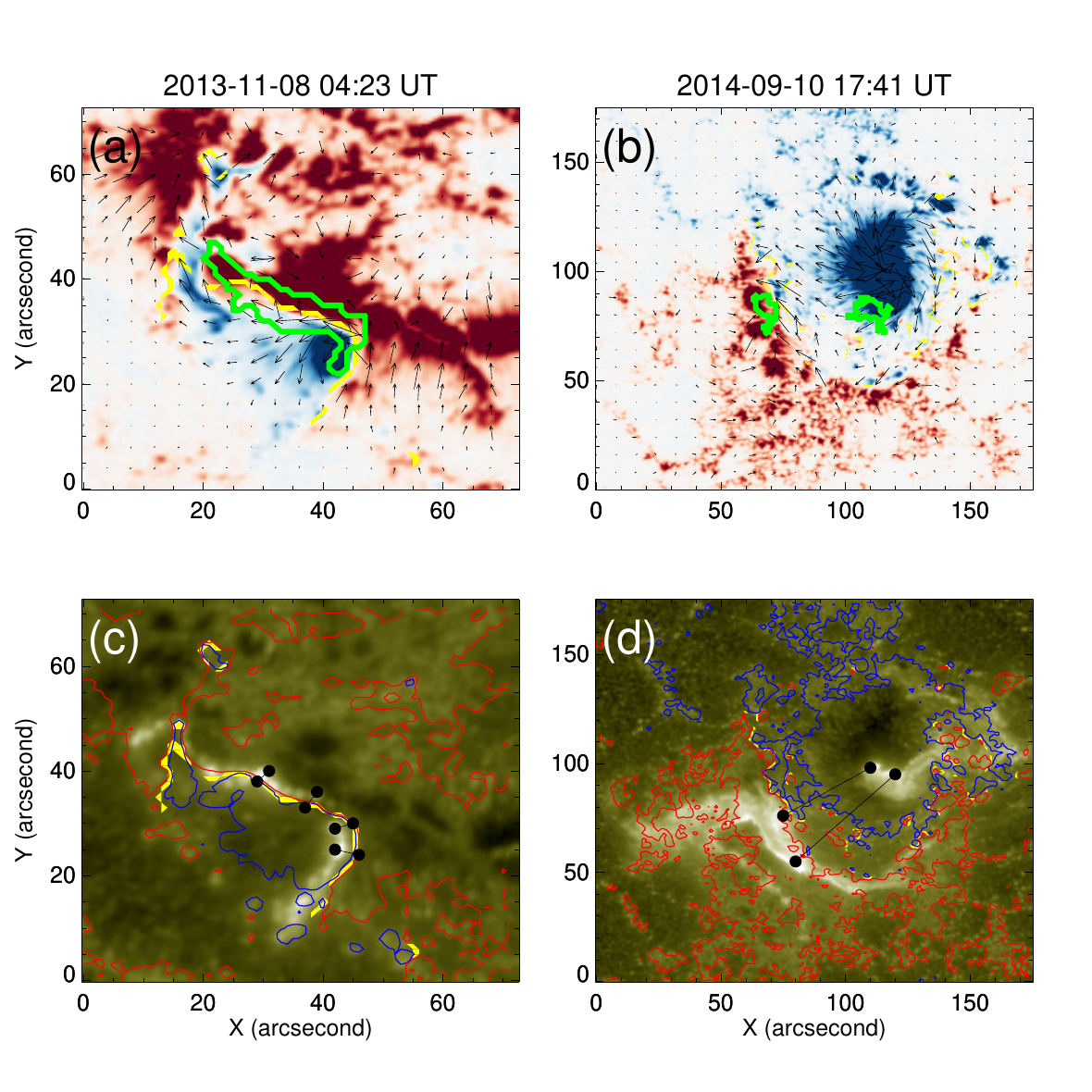}
\caption{Illustrations demonstrating the calculation of ribbon distances for two events. Panels a and b represent the photospheric magnetic field maps for the two events occurred on 2013-11-08 04:23 UT and 2014-09-10 17:41 UT respectively. The red/blue colors represent the positive/negative polarities of $B_r$ plotted within a range of $\pm 500$ Gauss. The yellow line is indicative of the polarity inversion line (PIL) and the green contour signifies the region where the significant change in Lorentz force is observed.} The black arrows represent the horizontal magnetic field lines. Panels c and d shows the flare ribbons as observed in the AIA 1600 \AA channels. The blue and red contours represent the overlaid positive and negative polarities of $B_r$ at levels $pm$ 100 Gauss. The solid black lines represent the ribbon distances.
\label{fig:aia_ribbons}
\end{figure*}

\begin{figure*}[!t]
    \centering
    \includegraphics[width=0.8\textwidth]{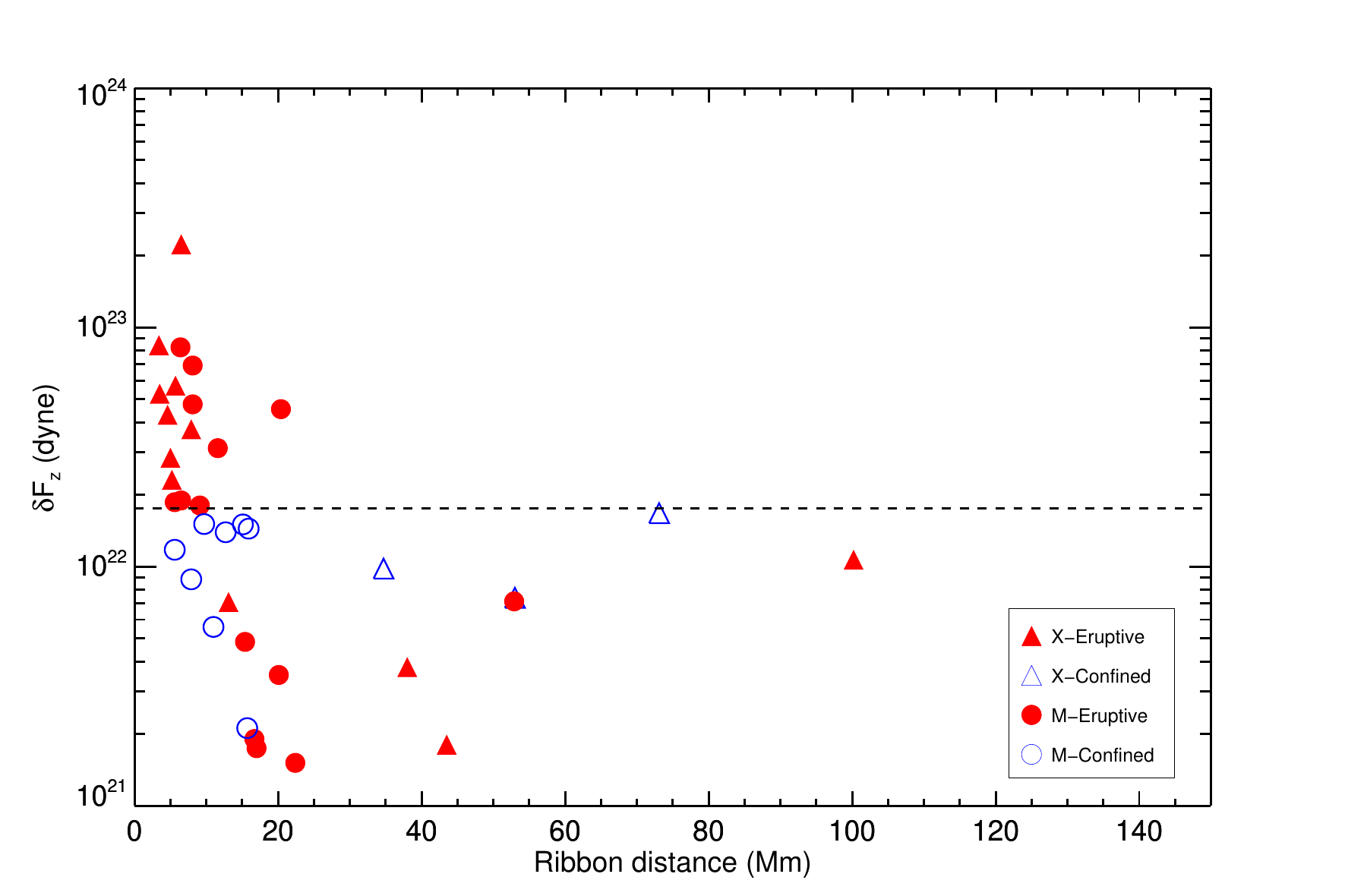}
    \caption{Scatter plot of vertical Lorentz force change vs ribbon distance. The values of Lorentz force change shown in the figure are estimated using the method based on the Lorentz force difference maps. Filled and empty symbols correspond to the eruptive and confined flares, respectively. The triangular and circular symbols imply X-class and M-class flares, respectively. The horizontal dashed line is drawn to illustrate the threshold value of change in Lorentz force. Few data points overlap each other.}
  \label{fig:ribbon_distance_plot}
\end{figure*}

\begin{figure*}[!ht]
\centering
    \includegraphics[width=0.8\textwidth]{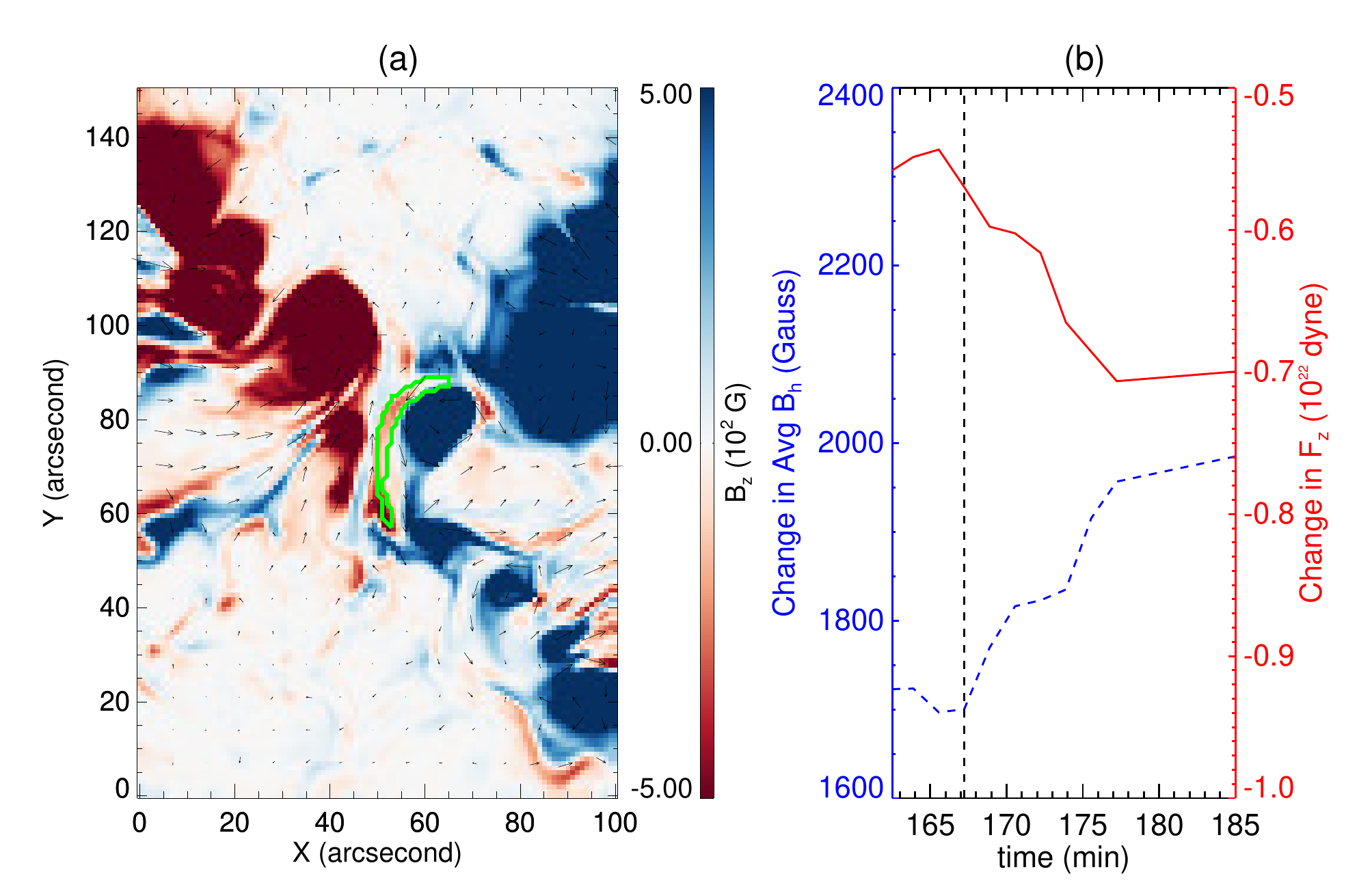}
    \includegraphics[width=0.8\textwidth]{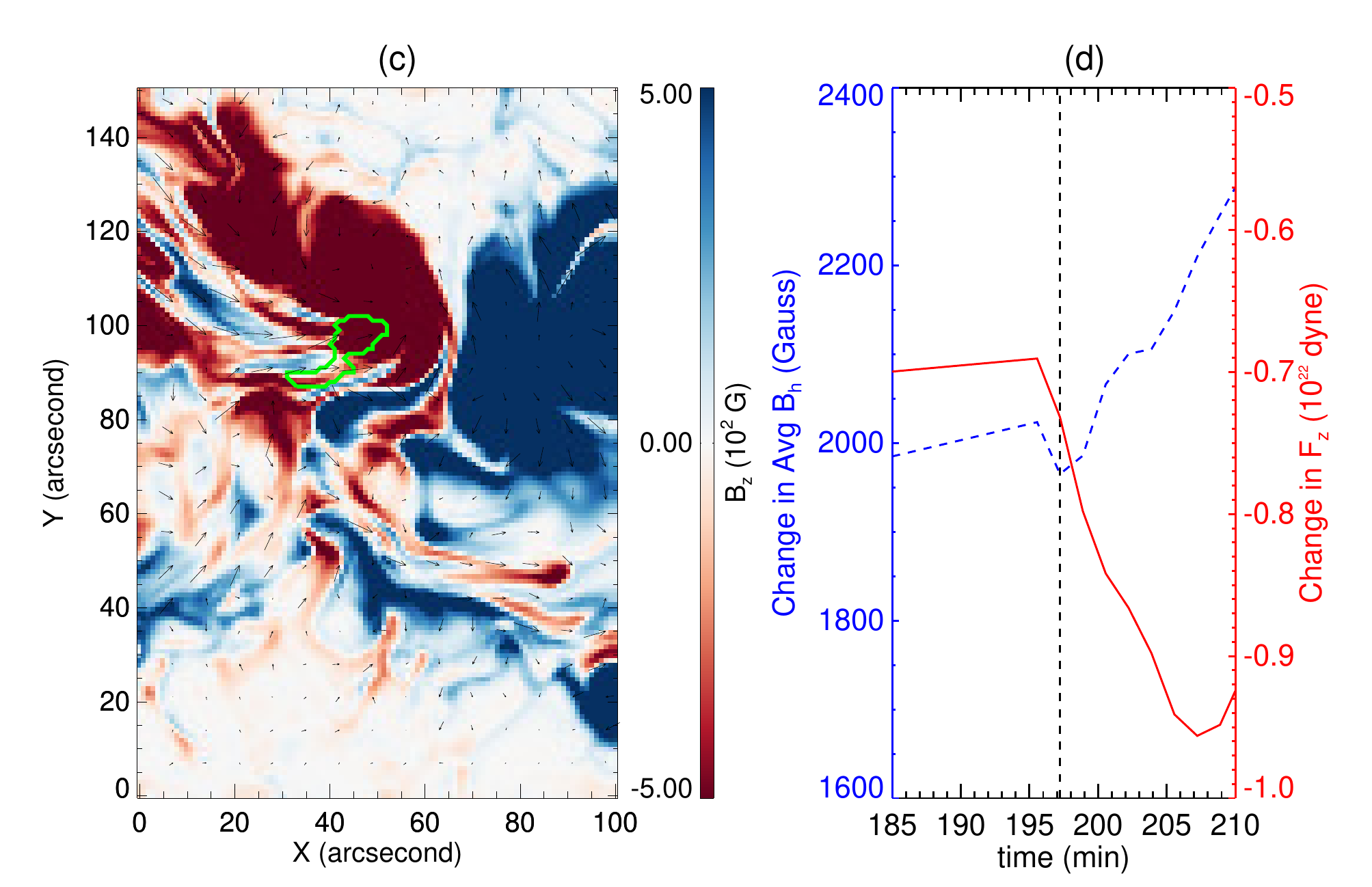}
\caption{(a) Illustration of the vertical magnetic field $B_z$ for the B-class synthetic flare events and (b) the corresponding temporal evolution of average horizontal magnetic fields (blue) and vertical Lorentz force (red). (c) and (d) Same as (a) and (b) but for the synthetic C-class flare. The green contours in (a) and (c) mark the region where significant change in Lorentz force occurs. The strength of the vertical magnetic field is represented by the colorbar. The dashed vertical black lines represent the flare time.}
\label{fig:sim_events}
\end{figure*}

\subsection{Statistics of $B_h$ and $F_z$ evolution}
In this subsection, we present the statistical properties of the average change $\delta B_h$ and the $\delta F_z$ for all the events listed in Table (\ref{tab:events}). The results show that the average $\delta B_h$ ranges from $15$ Gauss to $425$ Gauss, while the Lorentz force change varies from $1.5 \times 10^{21}$ dyne to about $22.3 \times 10^{22}$ dyne when the RoI was identified based on the change in vertical Lorentz force. Similarly, the average $\delta B_h$ ranges from $45$ Gauss to $630$ Gauss, while the Lorentz force change varies from $1.5 \times 10^{21}$ dyne to about $19.8 \times 10^{22}$ dyne when the RoI was identified based on the change in horizontal magnetic field. The data variations result from the adoption of different methods to identify the Region of Interest (RoI). This approach was employed to prevent any potential bias in the analysis. However, it is important to note that the results do not show significant differences and remain within the error limits.

Figure \ref{fig:bhcontour_plot}, \ref{fig:flcontour_plot} illustrates the change in average $B_h$ and total downward $F_z$ plotted against the flare strength for $B_h$ and $F_z$ contouring method respectively. The change in average $B_h$ does not exhibit statistically significant differences between eruptive and confined events, which is consistent with the findings of \cite{liu2022horizontal} and \cite{yadav2022statistical}.

However, the change in $F_z$ with flare strength clearly distinguishes between the two types of events. All confined flares have $\delta F_z < 1.8 \times 10^{22}$ dyne, whereas most eruptive flares show higher $\delta F_z$ values than the above mentioned limit. This suggests that the strength of flare associated $\delta F_z$ depends on whether the flare is eruptive or confined. The threshold limit can serve as a criterion for determining the presence of the associated CMEs based on these calculations. This reveals that while the change in average $B_h$ does not discriminate between eruptive and confined events, the change in total downward Lorentz force provides a clear distinction, indicating that magnetic imprints on the photosphere can be indicative of flare eruptivity.

It is worth noting that out of the 26 eruptive events, 9 of them exhibit a change of $F_z$ below the previously mentioned threshold limit. We present one such event in Figure \ref{fig:aia_ribbons}(b) \& (d). In Figure (b), the two green contours represent the regions where the most significant change in Lorentz force is observed. Interestingly, in this event, the strong ($>$100 Gauss) opposite polarity regions are not in close proximity to each other. This spatial arrangement seems to have an impact on the magnetic field dynamics. On the other hand, the rest of the 17 eruptive flares associated with $\delta F_z$ greater than the threshold value, are observed to occur where the regions of strong opposing polarities are close to each other (see an example of such event in Figure \ref{fig:aia_ribbons}(a) \& (c)).

This indicates that the morphology of the active region may be a contributing factor for the relatively smaller change in $F_z$ in our investigation. The spatial distribution of magnetic polarities within the active region appears to play a role in shaping the observed changes in magnetic fields and Lorentz forces during eruptive events.

This can be better understood with the help of a cartoon, as shown in the Figure \ref{fig:Cartoon}. The cartoon illustrates how the spatial distribution and arrangement of opposite polarity regions within an active region can influence the observed changes in magnetic fields and Lorentz force during flaring events. The solar flares are usually observed with two parallel ribbons, lying both sides of the PIL. If the two flare ribbons are thought of connected via newly reconnected semi-circular magnetic loops, then the distance between the two parallel ribbons would be proportional to the length of the loop and the reconnection height \citep{toriumi2017magnetic,Reep2017Dribbon}. Thus, a shorter ribbon distance would corresponds to smaller loop, whereas a longer ribbon distance corresponds to a larger loop in the solar atmosphere. For those events where the opposite polarity regions are in close proximity, the distance between the two parallel ribbons during the flare onset time is much shorter, as they form nearer to the PIL. This suggests that, in such cases, reconnection initiates at a lower altitude, resulting in a more significant impact on the photosphere characterized by larger changes in the Lorentz force \citep{liu2022horizontal,yadav2022statistical}. However, if the strong opposite polarity regions are not in close proximity to each other, the flare ribbons start a bit away from the PIL indicating that the reconnection begins at a higher height. As a result, the impact on the solar photosphere is less in this situation, which justifies a smaller change in Lorentz force that we observed.

This explanation is well consistent with the results shown in Figure \ref{fig:aia_ribbons} (c) \& (d), where the ribbon distance ($d_{ribbon}$) is estimated during the onset of the associated flares. As the flare ribbons mark the footpoints of the reconnecting magnetic loops, half of the distance ($d_{ribbon}/2$) between the two flare ribbons estimated during the onset time of the flare serves a proxy for the initial reconnection height in the solar corona. We use the observations from AIA 1600 \AA\ channel to identify the flare ribbons as shown in Figure \ref{fig:aia_ribbons}.

The method utilized for estimating the ribbon distance ($d_{ribbon}$) is presented with the example of two flares from our event list as depicted in Figure \ref{fig:aia_ribbons}. The first one was the X1.1 eruptive flare that occurred on 8 November 2013 (event-19) and the second one was X1.6 eruptive flare occurred on 10 September 2014 (event-25). Although both the flaring events were eruptive in nature, the change in Lorentz force associated with event-19 surpasses the critical threshold, whereas for event-25, it falls below the critical threshold. In order to understand the distinct morphological differences between this two events, we first identify the proximity of the polarity inversion line (PIL) by super-imposing the contours of $B_r$ at levels $\pm$ 100 Gauss onto the AIA1600 \AA\ images. Additionally, we apply the automated algorithm to identify the PIL (indicated by the yellow lines) as developed in \cite{sarkar2018comparative}. Figure \ref{fig:aia_ribbons} (c) depicts that the flare ribbons during the onset time of the flare for event-19, form very close to the PIL. Moreover, the associated HMI observations show that the opposite polarity regions of strong $B_r$ are closely located, forming a compact field region near the flaring PIL. As the flare ribbons at the either side of the PIL for event-19 does not form parallel to each other, we manually selected multiple points on the PIL from which we measured the shortest distance to the ribbon. Furthermore, taking an average of those multiple measurements and multiplying the mean distance with a factor of 2, we estimate the distance $d_{ribbon}$. In contrast to the event-19, the flare ribbons in event-25 form much away from the PIL and the opposite polarity regions of strong $B_r$ also locate away from the PIL, forming a dispersed field region at the flaring location (see Figure \ref{fig:aia_ribbons} (b) \& (d)). As the flare ribbons in event-25 form parallel to each other, we select points along the ribbons located either side of the PIL to estimate the average distance ($d_{ribbon}$) between the two ribbons. We apply the above mentioned method to all the events and list the estimated $d_{ribbon}$ in Table \ref{tab:events}.

The change in Lorentz force is plotted in Figure \ref{fig:ribbon_distance_plot} against $d_{ribbon}$. It is evident from the plot that the majority of eruptive events exhibit a ribbon separation smaller than $9$ Mm.\

\begin{figure}[!ht]
    \centering
    \includegraphics[width=\linewidth]{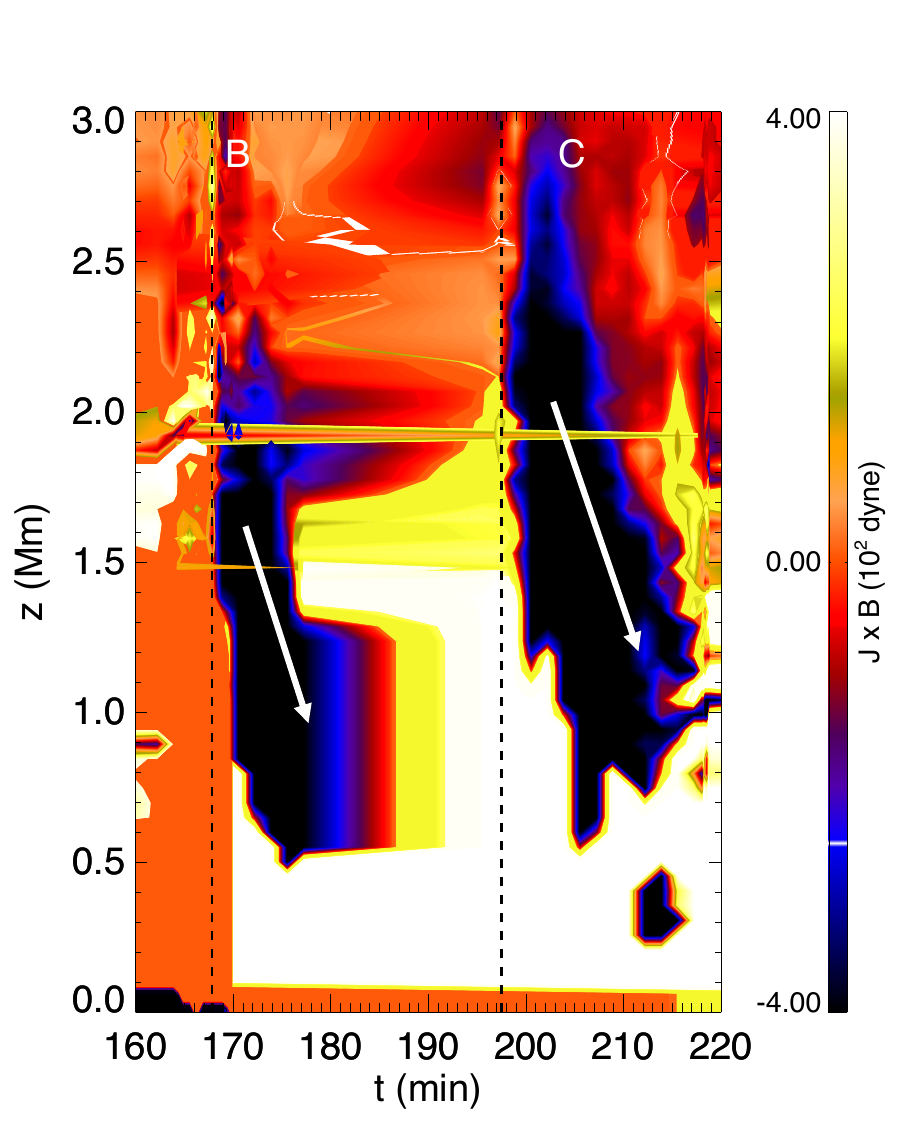}
    \caption{Height--time diagram of Lorentz force from the 3D MHD simulation to show its downward propagation. The dashed vertical lines represent the time of the two synthetic flares. The arrow is for guiding the eye towards the propagation direction of the Lorentz force.}
  \label{fig:lf_propagation}
\end{figure}

In contrast, those eruptive events that show a smaller change in the Lorentz force tend to have higher ribbon distances, typically exceeding $15$ Mm. This indicates a potential implication between the ribbon separation and the magnitude of the change in Lorentz force during eruptive events. Thus the ribbon separation could serve as an additional factor to consider when studying the magnetic imprints associated with the solar flares.
Overall, this helps to visualize how the specific morphology of the active region plays a crucial role in determining the magnitude of changes in the total downward Lorentz force ($F_z$), and this may be one of the factors contributing to the observed variations during such events.

\subsection{Downward propagation of the Lorentz force}
 We applied the same procedure as in observational data analysis to the B \& C class synthetic flare events and the similar profile of $B_h$ and $F_z$ are observed. Figure \ref{fig:sim_events} presents the results of our analysis performed on the simulation data. Utilizing our semi-automatic code, we are able to select the RoIs accurately (see the green contours in Figure \ref{fig:sim_events} [a] and [c]), which agrees with the flaring region identified by \cite{korsos2018weighted} based on the temperature anomaly. The dashed vertical lines in panels [b] and [d], represent the onset time of B and C-class flare. The change of $B_h$ in B and C-class flare is $60$ Gauss and $400$ Gauss, respectively, whereas, the change in $F_z$ is $0.15 \times 10^{22}$ dyne and $0.27 \times 10^{22}$ dyne respectively.

We have noticed a remarkable resemblance in the temporal evolution in the simulation with the observational data for the changes in horizontal magnetic field and vertical Lorentz force. While we cannot directly compare the simulation results to our observational data due to differences in flare class and size of active regions involved, the fact that both the horizontal magnetic field and Lorentz force display similar variations indicates that the underlying physics governing the changes in magnetic properties remains consistent across different flare classes.

Our simulation allows us to calculate the total Lorentz force, defined by $\mathbf{J}\times \mathbf{B}$, in a 3D setup. In contrast to observational data, where magnetic field components are only available at the solar surface, our simulation provides all physical variables defined at all heights from the photosphere to the corona. We plotted the height time plot of Lorentz force averaged over the horizontal plane for the two synthetic events as shown in Figure \ref{fig:lf_propagation} which agrees with the eruption time mentioned in \cite{korsos2018weighted}. This comes out to be of the order of $10^2$ dyne whereas the volume integral of average $\mathbf{J}\times \mathbf{B}$ over the domain comes out to the order of $10^{22}$ dyne which agrees with the observational results. We observed that the Lorentz force propagates towards the photosphere from the reconnection site similar to \cite{barczynski2019flare}. The average propagation speed is 2.4\,km$s^{-1}$ and  2.3\,km$s^{-1}$ for simulated B and C-class flares, respectively. Therefore, we argue that the Lorentz force from the reconnection site (marked 'X' in Figure \ref{fig:Cartoon}) propagates downward towards the photosphere, resulting in changes of $F_z$ and $B_h$ as estimated from the HMI magnetogram.

 Our analysis of the synthetic flare events provides further support for the importance of the downward propagation of the Lorentz force and its influence on the photosphere, which is consistent with both previous simulation studies and the observational data.

\section{Summary}
We present a statistical analysis of the flare-associated changes in the photospheric magnetic field close to the neutral line of the flaring region during the 37 flare events stronger than GOES M5 class flare from 26 different ARs. We used a semi-automatic technique to pick the sub-region close to the polarity inversion line where the significant changes in Lorentz force occur during the flare. We investigate whether the flare associated changes in Lorentz force and the photospheric magnetic field have any dependency on the confined or eruptive nature of a flare.

We have found a consistent pattern in the variation of the change in horizontal photospheric magnetic field ($B_h$) and change in vertical Lorentz force ($F_z$) for each event using two different methods to identify the most significant field change. The mean $B_h$ appears to increase in every case, showing abrupt enhancement in the temporal evolution. The observed increase in the horizontal magnetic field, $B_h$, can be attributed to the coronal implosion conjecture, as proposed by \cite{hundson2000implosions} or reconnection-driven contraction in post-flare loops, as shown by \cite{barczynski2019flare} using a zero-$\beta$ approximation MHD simulation.

We also observed a significant, abrupt, permanent downward change in vertical Lorentz force during each flare, demonstrating an abrupt change in the temporal evolution, which is a common feature in large flares \citep{sun2012evolution,wang2012response}. To understand the behavior, we compared the results of the MHD simulation of a solar flare and found that, the Lorentz force $(J \times B)$ propagates downward towards the photosphere over time, leading to the observed change in vertical Lorentz force. We observed similar temporal evolution profiles for the average $\delta B_h$ and the total $\delta F_z$ for both of these synthetic flares. This consistency in the temporal evolution patterns of $\delta B_h$ and $\delta F_z$ further supports the robustness and validity of our analysis method for studying flares, both in observational data and in simulated events.

Our conclusion regarding the distinction between eruptive and confined flares is based on a comparison of the $\delta F_z$ (change in vertical Lorentz force) for both types of flares. Eruptive flares were found to leave a significant magnetic imprint on the solar photosphere. Conversely, confined flares exhibited comparatively smaller changes in the photospheric magnetic field \citep{sun2012evolution,sarkar2018comparative}. This observation is in agreement with the flare-related momentum balance condition, which suggests that the Lorentz-force impulse is directly related to the associated CME momentum \citep{fisher2012global,wang2012relationship}. This is strongly supported by our findings from the analysis of the selected 26 eruptive and 11 non-eruptive flares in this work. The significant change in the photospheric magnetic field for eruptive flares and the comparatively smaller change for confined flares support the notion that the Lorentz-force impulse is linked to the presence and strength of associated CMEs.

However, to distinguish between the two types of flares, our analysis offers an upper threshold value of vertical Lorentz force change. Despite nine eruptive events having Lorentz force change below our threshold, we did not witness any confined events with Lorentz force larger than $1.8 \times 10^{22}$ dyne. Using 21 flaring episodes, between August 2010 and November 2015, \cite{vasantharaju2022magimprint} suggested that the strength of the magnetic imprint is independent of whether the flare is eruptive or not. On the contrary, our findings indicate that the mean $B_h$ change makes no distinction, whereas the change in vertical Lorentz force provides a clear differentiation between confined and eruptive flares. This difference is likely due to the inclusion of the term $\delta B_z^2$ when computing $\delta F_z$, identifying the RoI using an improved algorithm and inclusion of larger sample of flaring events in our study with energies equal to or exceeding those of M-class flares.

For eruptive events with Lorentz force change below the threshold, we noticed a significantly higher separation distance between the parallel flare ribbons when they form during the onset time of the flare. The source location of those events also displays distinct morphology, as the distance between the two strongly opposing magnetic-polarity regions at the flaring location is observed to be comparatively larger.  This larger separation is due to a higher reconnection height at the start of the flare as compared to the other eruptive events, leading to weaker Lorentz force change in the photosphere. Overall, our analysis reveals that the change in the vertical Lorentz force plays a crucial role in distinguishing confined and eruptive flares. The observed differences are influenced by factors such as the CME association and the separation distance of the parallel flare ribbons at the time of flare onset.

In this paper, we have examined the effects of major flares on fields near neutral lines. The present analysis is a step forward to distinguish the eruptive and confined flares in terms of the change in the vertical Lorentz force. Although a clear distinction between confined and eruptive events has been seen in this sample, it will be instructive to look at a larger sample of events and their corresponding vector field measurements from HMI and their associations with the CMEs.

We thank the referee for helpful comments that improved the quality of this manuscript. SSM acknowledges Ayan Ghosh for his contribution in illustrating the sketch in Figure 6. RS acknowledges support from the project EFESIS (Exploring the Formation, Evolution and Space-weather Impact of Sheath-regions), under the Academy of Finland Grant 350015. Additionally, authors thank the NASA SDO team  for providing valuable HMI and AIA data. SDO is a mission under NASA's Living with a Star program. The work was supported by Indo-US Science and Technology Forum (IUSSTF/JC-113/2019). Authors acknowledge NOVA HPC and Delphinus server of IIA used to perform most of the analysis. Open access is funded by Helsinki University Library.

\startlongtable
\begin{longrotatetable}
\begin{deluxetable*}{@{\extracolsep{4pt}} c c c c c c c c c c c c c}
\tablecolumns{12}
\centerwidetable
\centering
\tablecaption{List of 37 flare events from 27 active regions and their associated magnetic properties}
\label{tab:events}

\tablehead{
No & Date       & Peak Time & NOAA AR & Position & Flare Class & Type \tablenotemark{1} & \multicolumn{2}{c}{$F_z$ Contour} & \multicolumn{2}{c}{$B_h$ Contour} & Ribon distance \\ \cline{8-9} \cline{10-11}
   &            &           &         &          &             &      & $\Delta B_h$ (Gauss) \tablenotemark{2}         & $\Delta F_z$ (dyne) \tablenotemark{3}        & $\Delta B_h$ (Gauss) \tablenotemark{2}         & $\Delta F_z$ (dyne) \tablenotemark{3}        &     (Mm)
}

\startdata
1  & 2011-02-13 & 17:38:00  & 11158   & S20E04   & M6.6        & E    & $293.7 \pm 67.1$       & $1.80 \pm 0.32$         & $275.6 \pm 63.1$            & $1.74 \pm 0.33$          & $9.1$      \\
2  & 2011-02-15 & 01:56:00  & 11158   & S20W10   & X2.2        & E    & $192.5 \pm 25.5$        & $3.75 \pm 0.87$         & $197.8 \pm 29.4$            & $3.10 \pm 0.88$           & $7.9$      \\
3  & 2011-03-09 & 23:23:00  & 11166   & N08W09   & X1.5        & C    & $293.7 \pm 44.1$        & $0.98 \pm 0.27$          & $312.1 \pm 45.9$            & $0.93 \pm 0.22$          & $34.7$     \\
4  & 2011-07-30 & 02:09:00  & 11261   & S20W10   & M9.3        & C    & $287.5 \pm 61.1$        & $1.39 \pm 0.26$          & $469.7 \pm 67.3$            & $1.29 \pm 0.19$          & $12.7$     \\
5  & 2011-08-03 & 13:48:00  & 11261   & N16W30   & M6.0        & E    & $341.6 \pm 23.4$        & $3.12 \pm 0.18$          & $304.5 \pm 24.3$            & $2.87 \pm 0.21$          & $11.6$      \\
6  & 2011-09-06 & 01:50:00  & 11283   & N14W07   & M5.3        & E    & $222.5 \pm 39.3$        & $1.85 \pm 0.31$          & $208.0 \pm 34.5$            & $2.24 \pm 0.47$          & $5.6$      \\
7  & 2011-09-06 & 22:20:00  & 11283   & N14W18   & X2.1        & E    & $370.4 \pm 24.2$        & $2.30 \pm 0.13$          & $361.2 \pm 25.3$            & $2.14 \pm 0.11$          & $5.2$      \\
8  & 2011-10-02 & 00:50:00  & 11305   & N12W26   & M3.9        & C    & $190.0 \pm 47.1$        & $0.21 \pm 0.06$          & $187.5 \pm 40.7$          & $0.21 \pm 0.10$          & $15.7$     \\
9  & 2012-01-23 & 03:59:00  & 11402   & N28W21   & M8.7        & E    & $316.6 \pm 53.2$        & $0.71 \pm 0.09$         & $289.7 \pm 55.4$            & $0.62 \pm 0.12$         & $52.9$     \\
10 & 2012-03-07 & 00:24:00  & 11429   & N17E31   & X5.4        & E    & $215.0 \pm 19.4$        & $8.42 \pm 0.90$          & $239.5 \pm 21.0$          & $8.56 \pm 0.96$          & $3.4$      \\
11 & 2012-03-09 & 03:53:00  & 11429   & N15W03   & M6.3        & E    & $233.3 \pm 19.6$        & $1.89 \pm 0.09$          & $233.6 \pm 22.1$            & $1.72 \pm 0.17$          & $6.5$      \\
12 & 2012-07-02 & 10:52:00  & 11515   & S17E08   & M5.6        & E    & $263.3 \pm 61.5$        & $0.35 \pm 0.11$          & $235.2 \pm 74.9$            & $0.20 \pm 0.05$           & $20.1$     \\
13 & 2012-07-12 & 16:49:00  & 11520   & S15W01   & X1.4        & E    & $225.0 \pm 33.8$        & $0.38 \pm 0.05$          & $225.1 \pm 33.8$            & $0.41 \pm 0.05$          & $38.0$     \\
14 & 2013-04-11 & 07:16:00  & 11719   & N09E12   & M6.5        & E    & $385.0 \pm 116.6$        & $0.15 \pm 0.02$          & $235.3 \pm 85.0$            & $0.15 \pm 0.01$          & $22.4$     \\
15 & 2013-10-24 & 00:30:00  & 11877   & S09E10   & M9.3        & E    & $264.0 \pm 94.0$        & $0.17 \pm 0.04$          & $630.2 \pm 196.7$            & $0.25 \pm 0.07$          & $17.0$     \\
16 & 2013-11-01 & 19:53:00  & 11884   & S12E01   & M6.3        & C    & $338.3 \pm 15.9$        & $1.50 \pm 0.11$          & $308.3 \pm 22.2$            & $1.43 \pm 0.14$          & $9.7$      \\
17 & 2013-11-03 & 05:22:00  & 11884   & S12W17   & M4.9        & C    & $335.0 \pm 44.1$       & $0.88 \pm 0.10$          & $287.9 \pm 45.6$            & $0.87 \pm 0.20$         & $7.9$      \\
18 & 2013-11-05 & 22:12:00  & 11890   & S12E44   & X3.3        & E    & $379.1 \pm 111.3$        & $2.85 \pm 0.98$          & $450.4 \pm 78.0$            & $2.82 \pm 0.98$          & $5.0$      \\
19 & 2013-11-08 & 04:26:00  & 11890   & S12E13   & X1.1        & E    & $425.0 \pm 22.9$        & $5.28 \pm 0.31$          & $384.3 \pm 23.2$            & $5.18 \pm 0.33$          & $3.5$      \\
20 & 2014-01-07 & 18:32:00  & 11944   & S15W11   & X1.2        & E    & $90.6 \pm 27.4$         & $1.06 \pm 0.16$          & $106.6 \pm 69.5$            & $0.6 \pm 0.12$          & $100.2$    \\
21 & 2014-02-02 & 09:31:00  & 11967   & S10E13   & M4.4        & C    & $286.5 \pm 215.1$       & $1.44 \pm 1.04$          & $218.7 \pm 279.5$        & $0.96 \pm 0.48$           & $15.9$     \\
22 & 2014-02-04 & 04:00:00  & 11967   & S14W06   & M5.2        & C    & $166.6 \pm 114.9$        & $1.50 \pm 0.62$         &$229.2 \pm 105.2$        & $1.27 \pm 0.32$           & $15.1$     \\
23 & 2014-03-29 & 17:48:00  & 12017   & N10W32   & X1.1        & E    & $402.0 \pm 79.0$        & $0.71 \pm  0.21$         & $230.9 \pm 52.6$            & $1.42 \pm 0.40$          & $13.1$     \\
24 & 2014-04-18 & 13:03:00  & 12036   & S20W34   & M7.3        & E    & $161.6 \pm 43.4$        & $0.19 \pm 0.04$          & $151.9 \pm 20.5$            & $0.36 \pm 0.13$           & $16.7$     \\
25 & 2014-09-10 & 17:45:00  & 12158   & N11E05   & X1.6        & E    & $122.0 \pm 36.0$            & $0.18 \pm 0.06$          & $141.6 \pm 51.4$            & $0.19 \pm 0.06$          & $43.5$     \\
26 & 2014-10-22 & 14:28:00  & 12192   & S14E13   & X1.6        & C    & $163.3 \pm 40.5$         & $0.74 \pm 0.23$         & $175.0 \pm 37.7$            & $0.79 \pm 0.18$          & $53.0$     \\
27 & 2014-10-24 & 21:41:00  & 12192   & S22W21   & X3.1        & C    & $179.1 \pm 53.5$        &  $1.67 \pm 0.43$          & $161.2 \pm 16.2$            & $0.94 \pm 0.21$         & $73.1$     \\
28 & 2014-11-07 & 17:26:00  & 12205   & N17E40   & X1.6        & E    & $297.5 \pm 68.8$        & $4.31 \pm 0.51$          & $342.1 \pm 70.1$            & $3.86 \pm 0.52$          & $4.6$      \\
29 & 2014-12-04 & 18:25:00  & 12222   & S20W31   & M6.1        & C    & $383.0 \pm 102.9$        & $0.55 \pm 0.12$          & $391.3 \pm $37.0            & $0.58 \pm 0.06$          & $11.0$     \\
30 & 2014-12-17 & 04:51:00  & 12242   & S18E08   & M8.7        & E    & $142.9 \pm 9.7$        & $4.55 \pm 0.62$          & $177.5 \pm 6.4$            & $4.50 \pm 0.70$           & $20.4$     \\
31 & 2014-12-18 & 21:58:00  & 12241   & S11E15   & M6.9        & E    & $165.0 \pm 8.2$        & $4.76 \pm 0.26$          & $172.5 \pm 7.8$            & $4.62 \pm 0.33$          & $8.1$      \\
32 & 2014-12-20 & 00:28:00  & 12242   & S19W29   & X1.8        & E    & $275.0 \pm 20.1$        & $5.70 \pm 0.72$         & $245.8 \pm 10.7$            & $6.63 \pm 0.95$          & $5.7$      \\
33 & 2015-03-12 & 14:08:00  & 12297   & S15E06   & M4.2        & C    & $243.0 \pm 38.7$        & $1.17 \pm 0.19$          & $217.7 \pm 28.3$            & $1.32 \pm 0.23$          & $5.6$      \\
34 & 2015-06-22 & 18:23:00  & 12371   & N13W06   & M6.5        & E    & $172.5 \pm 6.6$        & $8.25 \pm 0.43$          & $172.9 \pm 7.6$            & $7.32 \pm 0.44$          & $6.4$      \\
35 & 2015-06-25 & 08:16:00  & 12371   & N12W40   & M7.9        & E    & $220.8 \pm 10.5$        & $6.92 \pm 0.44$          & $150.8 \pm 8.9$            & $6.90 \pm 0.56$           & $8.1$      \\
36 & 2017-09-04 & 20:33:00  & 12673   & S10W11   & M5.5        & E    & $145.8 \pm 109.6$        & $0.48 \pm 0.29$          & $157.9 \pm 106.2$            & $0.36 \pm 0.21$          & $15.4$     \\
37 & 2017-09-06 & 12:02:00  & 12673   & S09W34   & X9.3        & E    & $383.3 \pm 98.19$        & $22.25 \pm 5.39$         & $437.2 \pm 133.3$            & $19.75 \pm 8.77$         & $6.5$     \\
\enddata

\tablenotetext{1}{Eruptive (E) and Confined (C) flare}
\tablenotetext{2}{Average change of horizontal magnetic field ($B_h$)}
\tablenotetext{3}{Change in vertical Lorentz force ($F_z$)}

\end{deluxetable*}
\end{longrotatetable}

\bibliographystyle{aasjournal} 
\bibliography{references}{} 

\end{document}